\documentclass[letterpaper, 10 pt, conference]{ieeeconf}  

\IEEEoverridecommandlockouts                              

\usepackage{amsfonts}
\usepackage{mathrsfs}  
\usepackage{times}
\usepackage{bm}
\usepackage[dvipsnames]{xcolor}
\usepackage{amsmath,amssymb,latexsym}
\usepackage{graphicx}
\usepackage{multirow}
\usepackage{hyperref}
\usepackage{booktabs, makecell, tabularx, tabulary}

\usepackage{dsfont}
\usepackage{epstopdf}
\usepackage{empheq}
\usepackage{caption}
 \usepackage{dsfont}
\usepackage{subcaption}
\usepackage{algorithm}
\usepackage{algpseudocode}
\usepackage{cite}

\newtheorem{definition}{Definition}

\newtheorem{remark}{Remark}

\newtheorem{theorem}{Theorem} 

\title{\LARGE \bf
Neural Distributed Controllers with Port-Hamiltonian Structures
}

\author{Muhammad Zakwan and Giancarlo Ferrari-Trecate
\thanks{This research is supported by the Swiss National Science Foundation under the NCCR Automation (grant agreement 51NF40 180545).}
\thanks{Authors are with the Institute of Mechanical Engineering, Ecole Polytechnique Fédérale de Lausanne (EPFL), CH-1015 Lausanne, Switzerland, email: \{muhammad.zakwan, giancarlo.ferraritrecate\}@epfl.ch}%
}

\begin{document}

\maketitle
\thispagestyle{empty}
\pagestyle{empty}

\begin{abstract}

Controlling large-scale cyber-physical systems necessitates optimal distributed policies, relying solely on local real-time data and limited communication with neighboring agents. 
However, finding optimal controllers remains challenging, even in seemingly simple scenarios. Parameterizing these policies using Neural Networks (NNs) can deliver good performance, but their sensitivity to small input changes can destabilize the closed-loop system. This paper addresses this issue for a network of nonlinear dissipative systems. Specifically, we leverage well-established port-Hamiltonian structures to characterize deep distributed control policies with closed-loop stability guarantees and a finite $\mathcal{L}_2$ gain, regardless of specific NN parameters. This eliminates the need to constrain the parameters during optimization and enables training with standard methods like stochastic gradient descent. 
A numerical study on the consensus control of Kuramoto oscillators demonstrates the effectiveness of the proposed controllers.

\end{abstract}

\section{Introduction}\label{sec:Intro}


Distributed control of \emph{large-scale} systems presents formidable challenges even in seemingly basic scenarios due to the constrained flow of information in real-time. 
Particularly, Witsenhausen’s counter-example \cite{Witsenhausen} demonstrated that, even under apparently ideal conditions (i.e. linear dynamics, quadratic loss, and Gaussian noise), a nonlinear distributed control policy can outperform the best linear one.
A stream of works such as \cite{lessard2011quadratic} has provided  necessary and sufficient condition, namely, Quadratic Invariance (QI), under which the distributed optimal controller is linear and corresponds to solving a convex optimization problem. 
However, real-world systems often violate QI assumptions due to inherent nonlinearities, non-convex control costs, or privacy limitations \cite{furieri2022distributed}. 
This necessitates venturing beyond linear control and exploring highly nonlinear distributed policies, such as those parametrized by Deep Neural Networks (DNNs).


DNNs have proved their capabilities in learning-enabled control \cite{brunke2022safe, tsukamoto2021contraction, dawson2022safe, furieri2022distributed}, and system identification \cite{nghiem2023physics, beintema2023deep, verhoek2022deep, revay2023recurrent, wang2022learning, scandella2023kernel, zakwan2022physically, di2023simba, di2023stable} of non-linear dynamical systems.
Indeed, NN control has been applied in diverse application domains, such as robotics \cite{brunke2022safe}, epidemic models \cite{asikis2022neural}, safe path planning \cite{dawson2022safe}, and Kuramoto oscillators \cite{bottcher2022ai}.    
Existing approaches to NN control design also include
 modelling the system under control as a NN from data  \cite{hewing2019cautious, armenio2019model, bonassi2022recurrent, terzi2021learning}.
Nevertheless, NNs can be susceptible to small changes in their inputs~\cite{zakwan2022robust}. 
This fragility can easily translate to neural control policies, potentially jeopardizing the stability of closed-loop systems \cite{revay2023recurrent}. 
Moreover, the large number of parameters and intricate interconnections within NNs make it difficult to verify them for safety certificates and use them in large-scale safety-critical applications~\cite{dawson2022safe}. 




In this paper, we leverage well-established port-Hamiltonian (pH) system  framework \cite{vanderSchaft2017} to parametrize distributed DNN control policies that are inherently endowed with a finite $\mathcal{L}_2$ gain regardless of the choice of  trainable parameters.
This results in an unconstrained optimization problem for DNN control design solvable using standard gradient-based methods such as stochastic gradient descent or its variants. 
This eliminates the need for computationally expensive approaches such as projection of weight matrices or constrained optimization techniques. 
Therefore, if the underlying system to be controlled is dissipative, the proposed DNN controllers guarantee closed-loop $\mathcal{L}_2$ stability both during and after the training.
Moreover, the learned distributed policies are optimal in the sense that they minimize an arbitrary nonlinear cost function over a finite horizon.

{\bf Related work} 
DNNs have shown promise in designing both static and dynamic distributed control policies for large-scale systems.
Notably, Graph Neural Networks (GNNs) have achieved impressive performance in applications like vehicle flocking and formation flying \cite{yang2021communication, tolstaya2020learning, khan2020graph, gama2021graph} thanks to their inherent scalable structure.
However, guaranteeing stability with general GNNs remains challenging,  often requiring restrictive assumptions like linear, open-loop stable system dynamics or sufficiently small Lipschitz constants \cite{gama2021graph}.
Such limitations can be impractical, potentially leading to system failures during the training phase before an optimal policy can be found \cite{brunke2021safe,cheng2019end}. 
Some remedies to rectify this problem include improving an initial known safe policy iteratively, while imposing the constraint that the initial \emph{region of attraction} does not shrink \cite{berkenkamp2018safe,richards2018lyapunov,koller2018learning}, and 
 leveraging integral quadratic constraints to enforce closed-loop stability of DNN controllers \cite{pauli2021offset}. 
 However, these approaches explicitly constrain the DNN weights, which may lead to infeasibility or hinder the closed-loop performance. 
 In contrast, our proposed method based on free parameterizations provides the same scalability as GNNs without imposing any constraints on the weight matrices to guarantee closed-loop stability. 
 Although previous work explored stable-by-design control based on mechanical energy conservation \cite{abdulkhader2021learning, duong2021hamiltonian}, these methods are limited to specific systems (e.g., SE(3) dynamics). On the other hand, our approach applies to a wider range of nonlinear systems. 
 

Recently, the notion of \emph{free parametrization} has emerged for learning-enabled control, where an NN controller is trained to ensure its weight matrices satisfy specific constraints (e.g., semi-definite programs) \emph{by design}. This allows us to bypass computationally expensive post-verification routines.
Based on this approach, the framework of Recurrent Equilibrium Networks (RENs) has been proposed in~\cite{revay2023recurrent}.
RENs are a class of neural discrete-time nonlinear dynamical models that ensure built-in stability and robustness. Notably, they possess the unique property of satisfying desired integral quadratic constraints regardless of their weight matrices. 
Despite their flexibility, RENs face several limitations. Firstly, they are restricted to capturing dynamics with quadratic storage functions, limiting their expressiveness for complex systems. Secondly, the free parameterization approach in \cite{revay2023recurrent, martinelli2023unconstrained}  cannot be directly applied to distributed systems where sparsity patterns in weight matrices are crucial. 
In contrast, our NN framework based on pH structures offers several advantages. It allows the use of arbitrary nonlinear storage functions to capture more complex dynamics. Additionally, it seamlessly integrate desired sparsity patterns into the weight matrices, enhancing flexibility without compromising stability and performance.
Building on RENs, the work \cite{massai2023unconstrained} presents an unconstrained parametrization approach for interconnecting subsystems with finite $\mathcal{L}_2$ gain, while guaranteeing the $\mathcal{L}_2$ stability of the overall system. However, this approach is limited to quadratic storage functions for subsystems, constraining the flexibility and generalization.
Although \cite{furieri2022distributed} presented a similar distributed NN framework based on pH systems that ensure passivity by design but not a finite $\mathcal{L}_2$ gain for the closed-loop system which is instead our main result.
Unlike passivity, a finite $\mathcal{L}_2$ gain guarantees stability even in the presence of  external disturbances or modeling errors, which is crucial for safe operation in uncertain environments \cite{vanderSchaft2017, khong2018converse}.

{\bf Contributions}
The main contributions of this paper can be summarized as follows: 
\begin{enumerate}
    \item We provide a free parametrization of distributed controllers that can seamlessly incorporate sparsity in their weight matrices and are inherently endowed with a finite $\mathcal{L}_2$ gain. 
     \item Our approach overcomes the limitation of being restricted to specific storage functions (e.g., quadratic), enabling its application to a broader range of nonlinear control problems.
    \item We demonstrate the efficacy of our learning-enabled controllers on a benchmark consensus problem for Kuramoto oscillators.
\end{enumerate}

{\bf Organization:} 
Following the Introduction, 
Section \ref{sec:prelims} provides some preliminaries and the problem formulation.
In Section \ref{sec:main_results}, we provide a  free parametrization of neural distributed controllers via Hamiltonian structures endowed with a finite $\mathcal{L}_2$ gain regardless of the choice of weight matrices.
Finally, the performance evaluation of our NN controllers is conducted in Section \ref{sec:experiments}, whereas Section \ref{sec:conclusion} concludes the paper.

{\bf Notation:} Let $\mathcal{G} = (\mathcal{V}, \mathcal{E})$ be an undirected graph with nodes $\mathcal{V} = \{ 1, \dots,N \}$ and edges $\mathcal{E}$, and let $\mathcal{P} \in \{0, 1\}^{N \times N}$ be the corresponding adjacency matrix.  
For a binary mask $\mathcal{M} \in \{0,1\}^{m \times n}$, we denote $\bm{W} \in \texttt{blkSparse}(\mathcal{M})$ if  $\bm{W}$ is a block matrix and $\mathcal{M}_{i,j} = 0 \Rightarrow \bm{W}_{i,j} = 0$. $\bm{A} = \texttt{blkdiag}(A_i)$ represents a block-diagonal matrix with matrices $A_0, A_1,\dots, A_i$  on the diagonal. 
The set of non-negative real numbers is $\mathbb{R}_+$
and the standard Euclidean $2$-norm is denoted by $\Vert \cdot \Vert$.
We represent the set of $\mathbb{R}^n$-valued Lebesgue square-integrable functions by $\mathcal{L}_2^n := \{ v : [0, \infty ) \rightarrow \mathbb{R}^n \vert \Vert v \Vert_2^2 := \int_0^\infty v(t)^\top v(t) dt < \infty\}$. We omit the dimension $n$ whenever it is clear from the context.
Then, for any two $v,w \in \mathcal{L}_2^n$, we denote the $\mathcal{L}_2^n$-inner product as $\langle v,w \rangle:= \int_0^\infty v(t)^\top w(t) dt$. Define the truncation operator $(P_\mathcal{T} v)(t) := v(t)$ for $t \leq \mathcal{T}$; $(P_\mathcal{T} v)(t) := 0$ for $t > \mathcal{T}$, and the extended function space $\mathcal{L}_{2e}^n := \{ v : [0, \infty ) \rightarrow \mathbb{R}^n \vert P_\mathcal{T} v \in \mathcal{L}_2, \forall \mathcal{T} \in [0,\infty)\}$. For any linear space $\mathcal{U}$ endowed with a norm $\Vert \cdot \Vert_{\mathcal{U}}$, we define a Banach space $\mathcal{L}_{2e}(\mathcal{U})$ that consists of all measurable functions $f:\mathbb{R}_+ \mapsto \mathcal{U}$ such that  $\int_0^\infty \Vert f(t) \Vert^2_{\mathcal{U}} dt < \infty$.
Throughout this paper a system will be specified
by an input–output map $\Sigma: \mathcal{L}_{2e}^m \rightarrow \mathcal{L}_{2e}^p$ satisfying $\Sigma(0) = 0$. Given two systems $\Sigma_1$ and $\Sigma_2$, the standard negative feedback configuration between them is denoted by $\Sigma_1 \Vert_f \Sigma_2$, see Fig. \ref{fig:small_gain_theorem}.
The maximal eigenvalue of a matrix ${A}$ is represented by $\bar{\lambda}({A})$.

\section{Preliminaries and Problem Formulation} \label{sec:prelims}

We consider a network $\Sigma_s$ of $N \in \mathbb{N}$ coupled nonlinear subsystems, each endowed with a feedback control policy. Let $\mathcal{G}_s = (\mathcal{V}_s, \mathcal{E}_s)$ represents the graph associated with the  couplings among subsystems, and let $\mathcal{P}_s$ be its corresponding adjacency matrix. Then, each subsystem is governed by
\begin{subequations}
  \begin{empheq}[left={\Sigma_{s,i}:\empheqlbrace\,}]{align}
     \dot{x}_i(t)&=f_i\left(x_i(t), \breve{x}_i(t), u_i(t)\right), \label{eq:system1}\\
    y_i(t) &= h_i(x_i(t)), \qquad \quad   \ \forall i \in \mathcal{V} \;, \label{eq:system2}
  \end{empheq}
\end{subequations}
where $x_i \in \mathcal{X}_i \subseteq \mathbb{R}^{n_i}$ is the state, $u_i \in \mathcal{U}_i \subseteq \mathbb{R}^{m_i}$ is the input, and $y_i \in \mathcal{Y}_i \subseteq \mathbb{R}^{p_i}$ is the output of the subsystem $\Sigma_{s,i}$, respectively. 
We define $\breve{x}_i$ as a stacked vector of states of the $1$-hop neighbors of subsystem $i$ according to $\mathcal{G}_s$, i.e. all subsystems that influence $x_i$.
We assume there exists a unique solution trajectory $x_i(\cdot)$ on the infinite time interval $[0, \infty)$ of the differential equations \eqref{eq:system1} for all initial conditions $x_i(0) \in \mathcal{X}_i$ and $u_i(\cdot) \in \mathcal{L}_{2e}(\mathcal{U}_i)$, and $y_i(\cdot) \in \mathcal{L}_{2e}(\mathcal{Y}_i)$.
We assume that the distributed system $\Sigma_s$ is dissipative according to the following definition.
\begin{definition}[Dissipativity, \cite{vanderSchaft2017}] \label{def:dissipativity}The subsystem $\Sigma_{s,i}$ is called dissipative w.r.t. to a supply rate $s_i : \mathcal{U}_i \times \mathcal{Y}_i \mapsto \mathbb{R}$, if there exists a smooth storage function $V_i:\mathcal{X}_i \mapsto \mathbb{R}_+$ such that
$$
\dot{V}_i({x_i}(t)) \leq s(u_i(t),y_i(t)), \quad  \forall t \in \mathbb{R}_+ \;,
$$ or equivalently,
$$ V_i({x}_i(\tau)) - V_i({x}_i(0)) \leq \int_0^\tau s_i(u_i(t),y_i(t)) dt \;,
$$
for every input signal ${u}_i(t) \in \mathcal{U}_i$, output signal $y_i(t) \in \mathcal{Y}_i$ and every $\tau \geq 0$. Moreover, the choice of supply rate leads to different notions of dissipativity, for instance, 
\begin{enumerate}
    \item[$\bullet$] if $p_i = m_i$, and $s_i(u_i(t),y_i(t)) = u_i(t)^\top y_i(t)$, then system $\Sigma_{s,i}$ is passive; 
    \item[$\bullet$]  if $p_i = m_i$, and $s_i(u_i(t),y_i(t)) = u_i(t)^\top y_i(t) + \epsilon \Vert y_i(t)\Vert$, then system $\Sigma_{s,i}$ is $\epsilon$-output strictly passive for $\epsilon>0$;
    \item[$\bullet$]  if $s_i(u_i(t),y_i(t)) = \gamma^2 \Vert u_i(t) \Vert + \Vert y_i(t) \Vert$, then system $\Sigma_{s,i}$ has finite $\mathcal{L}_2$ gain, i.e. $\Vert y_i(t) \Vert \leq \gamma \Vert u_i(t) \Vert + b$ for some  non-negative constants $\gamma, b$.   
\end{enumerate}

\end{definition}

Recall that $\epsilon$-output strict passivity also implies a finite $\mathcal{L}_2$ gain not larger than $1 / \epsilon$ \cite{vanderSchaft2017}. 
Note that the storage function ${V}(\cdot)$ can be interpreted as the stored ``energy" in the system w.r.t. a single point of neutral storage (minimum energy).

The distributed control of large-scale systems  presents a major challenge: local controllers at each subsystem $u_i(t)$ can only access real-time information from a limited set of neighbors, dictated by a communication network $\mathcal{G}_c$. This network is represented by an adjacency matrix $\mathcal{P}_c \in \{0, 1\}^{N \times N}$ where $\mathcal{P}_{c_{i,i}} = 1$ for every $i \in \mathcal{V}$.
\begin{figure}
    \centering
    \includegraphics[scale = 0.5]{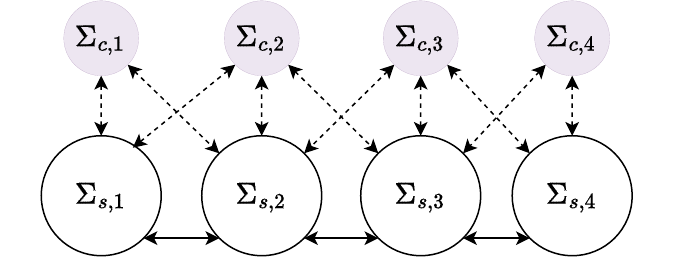}
    \caption{An example of a large-scale system $\Sigma_s$ and a distributed controller $\Sigma_c$ for $N = 4$.
    The solid lines represent interactions between the subsystems of $\Sigma_s$, and the dashed lines represent the flow of information between the system $\Sigma_s$ and the controller $\Sigma_c$.}
    \label{fig:distributed}
\end{figure}
 In this paper, our goal is to develop distributed dynamic feedback controller $\Sigma_c$ represented by the pairs $(\chi_i(\cdot),\pi_i(\cdot)), i \in \mathcal{V}$ defining local controllers 
\begin{equation} \label{eq:cont_org}
\Sigma_{c,i}: \begin{cases}
        \begin{aligned}
        \dot{\xi}_i(t) &= \chi_i(\xi_i(t), \breve{y}_i(t)), \quad \xi_i \in {\Xi}_i \subseteq \mathbb{R}^{q_i},\\
        u_i(t) &= \pi_i(\xi_i(t),\breve{y}_i(t)) \;,
    \end{aligned}
\end{cases}
\end{equation}
where $\xi_i$ is the state of $\Sigma_{c,i}$, and $\breve{y}_i(t)$ is a stacked vector of outputs from the neighbouring subsystems based on the communication graph $\mathcal{G}_c$. 
An example of a communication graph between the controller $\Sigma_c$ and the distributed systems $\Sigma_s$ is illustrated in Fig. \ref{fig:distributed}.
Moreover, the control policies parametrized by $\Sigma_c$ should be \emph{optimal} in the sense that they minimize an arbitrary  real-valued cost function 
\begin{equation} \label{eq:loss_function}
    c(\bm{x}(t),\bm{u}(t)) = \frac{1}{T} \int_0^T \ell(\bm{x}(t), \bm{u}(t)) dt
\end{equation}
for a finite horizon $T \in \mathbb{R}_+$, where $c$ is differentiable almost everywhere.  The bold-faced signals $\bm{x}(t) = [x_1^\top, \dots, {x}_N^\top]^\top, \bm{u}(t) = [u_1^\top, \dots, {u}_N^\top]^\top$ represent concatenated local states and local inputs, respectively. 
Finally, we assume that the set of tuples $\{ (\chi_i(\cdot), \pi_i(\cdot)) \}, i \in \mathcal{V}$ is parametrized by some NNs with trainable parameters $\theta_i \in \mathbb{R}^{d_i}$ for $i \in \mathcal{V}$. 
We define $\bm{\theta} = (\theta_1, \dots, \theta_N)$.

Besides designing optimal control policies, ensuring the stability of the  closed-loop $ \Sigma_s \Vert_f \Sigma_c$ formed by the distributed system $\Sigma_s$ and the NN controller $\Sigma_c$ is equally crucial. 
To address this issue, we focus on achieving $\mathcal{L}_2$ 
stability for the closed-loop system $\Sigma_s \Vert_f \Sigma_c$, leveraging the following results.
\begin{figure}
    \centering
    \includegraphics[width = 0.8\linewidth]{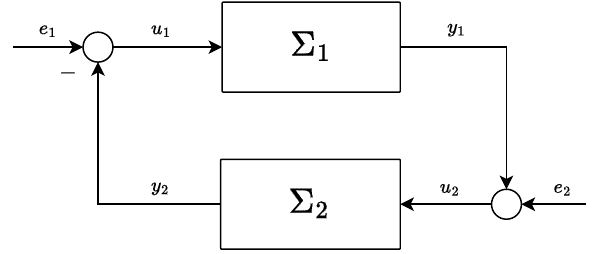}
    \caption{Standard feedback interconnection $\Sigma_1 \Vert_f \Sigma_2$.}
    \label{fig:small_gain_theorem}
\end{figure}
\begin{theorem}[\cite{vanderSchaft2017}] \label{thm:small_gain}
Consider the closed-loop system $\Sigma_1 \Vert_f \Sigma_2$ given in Fig. \ref{fig:small_gain_theorem}.   
\begin{enumerate}
    \item[$\bullet$] (small gain condition) Assume the existence of  the $\mathcal{L}_2$ gains $\mathcal{L}_2(\Sigma_1) \leq \gamma_1$, and $\mathcal{L}_2(\Sigma_2) \leq \gamma_2$. Then, the closed-loop system $\Sigma_1 \Vert_f \Sigma_2$ is stable with an $\mathcal{L}_2$ gain $ \gamma_1 . \gamma_2 < 1;$
\item[$\bullet$]  (strict output passivity) Assume that, for any $e_1 \in \mathcal{L}_{2e}(\mathcal{U}_1)$ and $e_2 = 0$,  $\Sigma_1:\mathcal{L}_{2e}(\mathcal{U}_1) \rightarrow \mathcal{L}_{2e}(\mathcal{Y}_1)$ is $\epsilon_1$-output strictly
passive, and $\Sigma_2:\mathcal{L}_{2e}(\mathcal{U}_2) \rightarrow \mathcal{L}_{2e}(\mathcal{Y}_2)$ is passive. Then,  $\Sigma_1 \Vert_f \Sigma_2$ for $e_2 = 0$ with input $e_1$ and output $y_1$ has  $\mathcal{L}_2$-gain $\leq 1 / \epsilon_1$. \hfill \QEDclosed
\end{enumerate}
\end{theorem}

Our goal is to train control policies for large-scale systems that address three key requirements:
\begin{enumerate}
    \item[i)] Limited information access: The policies must operate with restricted local information.
    \item[ii)] Optimal performance: They achieve optimal behavior by empirically minimizing an arbitrary user-defined  cost function.
    \item[iii)] Guaranteed stability: The closed-loop system is $\mathcal{L}_2$ stable.
\end{enumerate} 
This can be formulated as the following optimization program 
\begin{align}
\label{opt_prob}
     \min_{\bm{\theta}} \quad & \frac{1}{S} \sum_{k = 1}^{S}c(\bm{x},\bm{u};\bm{\theta}, \bm{x}_0^k) \\
    \text{s.t.} \quad &\text{system dynamics $\Sigma_s$}  \nonumber \\ 
      &\dot{\xi}_i(t) = \chi_i(\xi_i, \breve{y}_i(t), \theta_i), \label{eq:cont_1}\\
      &u_i(t) = \pi_i(\xi_i,\breve{y}_i(t),\theta_i), \quad \forall i \in \mathcal{V}, \label{eq:cont_2} \\ 
      &\eqref{eq:cont_1}-\eqref{eq:cont_2} \ \text{has a finite} \ \mathcal{L}_2 \ \text{gain}, \forall \bm{\theta} \in \mathbb{R}^{\bm{d}} \label{const:l2} \;,
\end{align}
where $S$ is the number of given initial conditions for $\Sigma_s$. The primary challenge of this optimization problem lies in finding the parameters $\bm{\theta}$ such that the distributed controller \eqref{eq:cont_1}-\eqref{eq:cont_2} has a finite $\mathcal{L}_2$ gain  without jeopardizing the standard NN training routines and without increasing the computational complexity. The next Section presents a novel method that tackles this challenge effectively.

\begin{remark}[Passivity by design]
While achieving passivity by design for the closed-loop system is considered in \cite{furieri2022distributed}, it may not always ensure stability, especially when controlled system interacts with a passive, but else completely unknown environment.
In fact, the converse of the passivity theorem tells us that the controlled system must be output strictly passive as seen from the interaction port of the controlled system with the environment~\cite{khong2018converse}.    
\end{remark}

\section{Neural $\mathcal{L}_2$-stable Hamiltonian Controllers}
\label{sec:main_results}
  
To address the $\mathcal{L}_2$ constraint \eqref{const:l2} in the optimization program \eqref{opt_prob}-\eqref{const:l2}, two common approaches \cite{dawson2022safe} involve constrained optimization or projection of $\bm{\theta}$ onto a set $\Theta_{\mathcal{L}_2}$ such that the NN controller $\Sigma_c$ has a finite $\mathcal{L}_2$ gain. 
However, these methods can be computationally burdensome, hence limiting the class of controllers that can be used. 
To circumvent this, we propose a \emph{free parametrization} approach  which involves designing a class of input-output operators that inherently possess a finite $\mathcal{L}_2$ gain for any choice of weight matrices $\bm{\theta}$. 
This enables us to seamlessly employ unconstrained optimization methods such as stochastic gradient descent and its variants to solve \eqref{opt_prob}-\eqref{const:l2}.  To achieve this, we leverage the well-established port-Hamiltonian framework \cite{vanderSchaft2017} to parametrize controllers $\Sigma_c$ with guaranteed finite $\mathcal{L}_2$ gains. 

Consider the following neural distributed pH controller with $N$ sub-controllers endowed with some trainable parameters
\begin{equation}
\Sigma_{pH}: \begin{cases}
\begin{aligned}
      \dot{\bm{\xi}}(t)& =\left[\bm{J}_{c}- (\alpha \bm{I} + \bm{\Lambda})\right] \frac{\partial H_{c}}{\partial \bm{\xi}}(\bm{\xi}, {\theta})+ \bm{G}_{c} \bm{y}(t) \\
    \bm{u}(t) & = \bm{G}_c^\top \frac{\partial H_{c}}{\partial \bm{\xi}}(\bm{\xi},\theta) \;,
\end{aligned}\label{eq.Controller}  
\end{cases}
\end{equation}
where $\bm{\xi} \in \Xi \subseteq \mathbb{R}^{n_c}, \bm{u}\in \mathcal{U}_c \subseteq \mathbb{R}^m, \bm{y} \in \mathcal{Y}_c \subseteq \mathbb{R}^m$ are  stacked vectors of NN controllers' states, outputs, and inputs, respectively. The interconnection matrix $\bm{J}_{c}=-\bm{J}_{c}^{\top} = \texttt{blkdiag}(J_i)$ is skew-symmetric.
The dissipation rate of $\Sigma_{pH}$ is determined by the damping matrix $\alpha \bm{I} + \bm{\Lambda}$, where $\alpha \in \mathbb{R}_+$, and $\bm{\Lambda} = \operatorname{diag}(e^{\bm{d}}) \in \mathbb{R}_+^n$ is a diagonal matrix for some free vector $\bm{d} \in \mathbb{R}^{n_c}$. 
The input matrix $\bm{G}_c = \texttt{blkSparse}(\mathcal{P}_c)$ is  full rank, where $\mathcal{P}_c$ is the underlying adjacency matrix of the communication topology.\footnote{
Decentralized control is achieved by setting $\bm{G}_c = \texttt{blkdiag}(G_i)$ in \eqref{eq.Controller}, making each sub-controller independent of the state if other subsystems or controllers.}
Moreover, the ``energy-like" Hamiltonian function $H_c: \mathbb{R}^{n_c} \rightarrow \mathbb{R}$ of  $\Sigma_{pH}$ is the algebraic sum of all $N$ controllers' energies $H_i$, i.e. $H_c (\bm{\xi}) = \sum_{i \in \mathcal{V}} H_i(\xi_i(t))$, where we assume that all functions $H_i$ are continuously differentiable and radially unbounded. 

Our main result concerns a free parametrization of the NN controller $\Sigma_{pH}$ that guarantees a finite $\mathcal{L}_2$ gain
 regardless of the choice of its trainable parameters. 
These parameters, collectively denoted by $\bm{\theta}$, encompass the weight matrices $\{ \bm{J}_c, \bm{\Lambda}, \bm{G}_c, \theta \}$.


\begin{theorem} \label{thm:finite_l2_gain}
Given a constant $\epsilon > 0$, let $\alpha = \epsilon \bar{\lambda}(\bm{G}_c \bm{G}_c^\top)$. Then, the NN controller $\Sigma_{pH}$
\begin{enumerate}
    \item[$\bullet$] is $\epsilon$-output strictly passive, and
    \item[$\bullet$] has a finite $\mathcal{L}_2$-gain $\leq 1/\epsilon$. \hfill \QEDclosed
\end{enumerate}
\end{theorem}

The detailed proof of Theorem \ref{thm:finite_l2_gain} is provided in the Appendix.
In simple words, Theorem \ref{thm:finite_l2_gain} implies that for any free choice of trainable parameters $\bm{\theta}$, one can always choose sufficiently large damping $\alpha$ such that the controller $\Sigma_c$ is $\epsilon$-output strictly passive, and consequently, the  map from $\bm{y}(t) \mapsto \bm{u}(t)$ has a finite $\mathcal{L}_2$ gain. 
Therefore, one can leverage Theorem \ref{thm:finite_l2_gain} and invoke Theorem \ref{thm:small_gain} to ensure closed-loop stability in cases where the system $\Sigma_s$ is passive (strict output passivity) or has a finite $\mathcal{L}_2$ gain (small gain condition).   

While we assume that the large-scale system $\Sigma_s$ is passive from $\bm{u}(t)$ to $\bm{y}(t)$ in our simplified setting, 
passivity truly holds only if all subsystems are individually passive (from $u_i(t)$ to $y_i(t)$) and the interconnection is power-conserving (e.g. skew-symmetric \cite{arcak2016networks}).
For deeper insights into preserving passivity and 
$\mathcal{L}_2$ gain in large-scale interconnected systems, refer to \cite{arcak2016networks}.
\begin{remark}[Selection of Hamiltonian]\label{remark_hamiltonian}
    We impose minimal restrictions on the Hamiltonian function $H_c(\bm{\xi}, \theta)$, i.e., differentiability and radial unboundedness. This flexibility allows for diverse choices, including simple quadratic functions, multi-layer perceptrons (MLPs), or even Hamiltonian deep neural networks (as in \cite{galimberti2023hamiltonian, zakwan2023universal} for representing $H_c$). Importantly, our results hold irrespective of the choice.
\end{remark}
\begin{remark}[Comparison with RENs] 
Unlike RENs \cite{revay2023recurrent}, our parametrization allows for incorporting diverse sparsity patterns within the weight matrices. 
Moreover, our method overcomes the limitation of RENs to use only quadratic storage functions. An in-depth comparison with the modelling capabilities of  RENs is difficult but, nevertheless, \eqref{eq.Controller} offers an alternative way to parametrize $\mathcal{L}_2$ operators.  
\end{remark}
\begin{remark}[Communication among sub-controllers]~Note that our framework can seamlessly incorporate communication graphs among the sub-controllers without loss of generality. 
In fact, the work \cite{furieri2022distributed} provides a systematic approach to interconnect sub-controllers while preserving the dissipativity of the closed-loop. 
For specific details and an example, we defer the readers to \cite[Theorem 3]{furieri2022distributed}.  
\end{remark}

{\bf Training of neural Hamiltonian controllers}
Having established the free parametrization of a class of distributed control policies that preserves closed-loop $\mathcal{L}_2$ stability by design, we now seek to solve the optimization problem \eqref{opt_prob}-\eqref{const:l2} by training a Neural ODE \cite{NeuralODEs} which is a NN model that extends standard layer-to-layer propagation to continuous-time dynamics.
In simple words, it is a non-linear ODE $\dot{x}(t) = {f}({x}(t),{\bm{\theta}(t)})$, where the vector field ${f}(\cdot)$ is represented by a NN. For details on Neural ODE implementations and its training, we refer the interested reader to \cite{NeuralODEs}. 
For our setting, the neural ODE  is the trainable closed-loop system $\Sigma_s \Vert_f \Sigma_{pH}$ with trainable parameters ${\bm{\theta}}$. 
We highlight that discretization during training might lead to sub-optimality. However, it does not compromise the closed-loop stability guarantees from Theorem \ref{thm:small_gain}. 
This is because the $\mathcal{L}_2$ gain of  continuous-time system $\Sigma_{pH}$ holds regardless of the weights, as long as $\alpha$ is chosen as in Theorem \ref{thm:finite_l2_gain}.



\section{Synchronization in Kuramoto Oscillators} \label{sec:experiments}

To showcase the efficacy of our NN control framework, we consider the problem of synchronization in Kuramoto oscillator model. Indeed, this problem is pervasive for investigating collective synchronous behaviors in several applications. For example, locking of circuit oscillators, frequency synchronization in power grids \cite{dorfler2014synchronization}, collective motion of self-propelled vehicles, and opinion synchronization in social networks. The interested readers are deferred to \cite{Wu2020collective} and references therein for more details.



A Kuramoto model consists of a population of $N$ oscillators whose dynamics are
\begin{equation}
\label{eq:kuramoto}
    \dot{\vartheta}_i = \omega_i + \frac{K u_i(t)}{N} \sum_{j = 1}^N \mathcal{P}_{ij} \sin(\vartheta_j - \vartheta_i), \ i \in \mathcal{V} \;,
\end{equation}
where $\vartheta_i$ is the phase and $\omega_i$ is the natural frequency of $i-$th oscillator, respectively.  Moreover, $K$ is the coupling strength and $\mathcal{P}_{ij}$ are the adjacency matrix components of the underlying (undirected) network. While the NN control of Kuramoto oscillators has been considered in \cite{bottcher2022ai}, their approach lacks closed-loop stability guarantees, which might lead to undesirable system behavior.
 \begin{figure}
    \centering
    \includegraphics[width=0.98\linewidth]{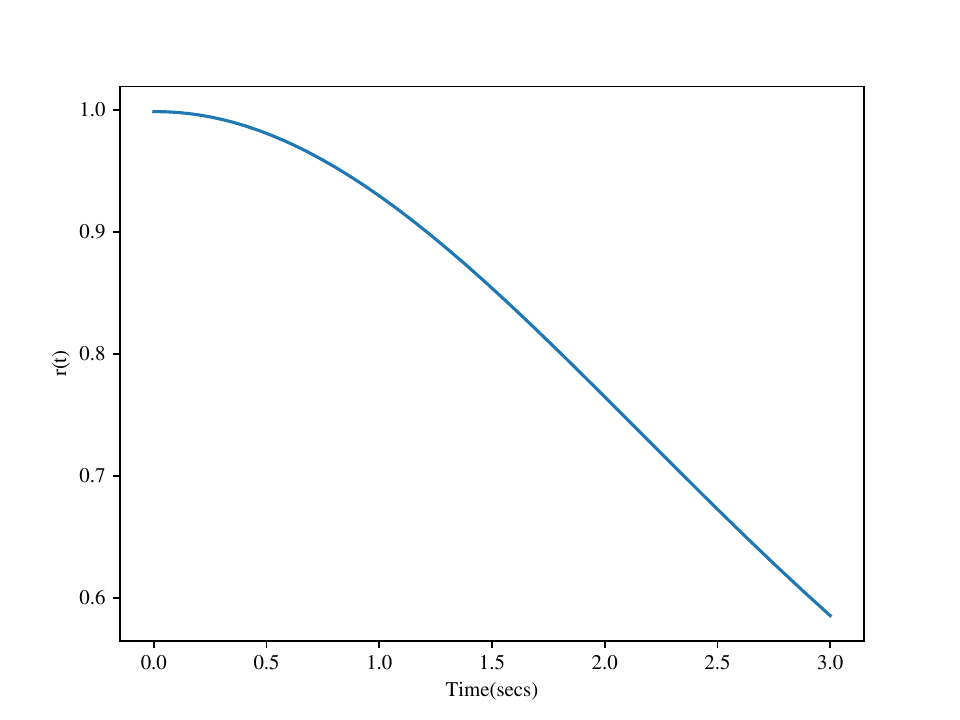}
    \caption{The consensus metric $r(t)$ for an uncontrolled network with a fully connected communication topology demonstrating unsynchronized behavior of the oscillators.}
    \label{fig:openloop}
\end{figure}

Let all the oscillators at $t = 0$ be initialized in the set $\mathcal{D} := \{\vartheta_i,\vartheta_j \ \text{s.t.} \ |\vartheta_i - \vartheta_j| < \frac{\pi}{2} \  \forall i,j = \mathcal{V}\}$. Then, by constructing the dynamics of the angular frequencies and differentiating the Kuramoto model \eqref{eq:kuramoto}, one obtains
\begin{equation} \label{eq:second_dynamics}
    \ddot{\vartheta_i} = \frac{K u_i(t)}{N}\sum_{j = 1}^N \mathcal{P}_{ij} \cos(\vartheta_j - \vartheta_i) (\dot{\vartheta}_j - \dot{\vartheta}_i) \;.
\end{equation}
By the change of variables $\dot{\vartheta}_i = x_i$ in \eqref{eq:second_dynamics}, we have 
\begin{align} 
\label{eq:passive_kuramoto}
     \dot{x}_i &= \nu, \ \nu = \frac{K u_i(t)}{N}\sum_{j = 1}^N \mathcal{P}_{ij} g_{ji} (x_j - x_i) \;, \\
     y_i &= x_i \;, \nonumber
\end{align}
  where $g_{ji} = \cos(\vartheta_j - \vartheta_i) $. 
  One can show that the system \eqref{eq:passive_kuramoto} is passive w.r.t. the storage function $V(\bm{x}) = \frac{1}{2}\bm{x}^\top \bm{x}$, where $\bm{x} = [x_1, \ \dots \, x_N]^\top$ is the concatenated vector representing the angular rates of the oscillators \cite{chopra2006passivity}.
  Our goal is the phase-synchronization of \eqref{eq:passive_kuramoto}  at some final time $T > 0$, that is,
  \begin{equation*}
      |x_i(T) - x_j(T)| = 0 \ \text{for} \ \mathcal{P}_{i,j} = 1\;,
  \end{equation*}
 while guaranteeing the closed-loop stability both during and after the training. 
 Thus, our objective can be formulated as the following optimization problem
\begin{align*}
     \min_{\bm{\theta}} c(\bm{x}(t),& \bm{u}(t); \bm{\theta}, \bm{x}_0) \  \text{s.t.} \ \text{the closed-loop is $\mathcal{L}_2$ stable} \nonumber \\
    c(\bm{x}(t),\bm{u}(t), \bm{x}_0) &= \frac{1}{2} \int_{0}^T \sum_{i,j} \mathcal{P}_{ij} \sin^2(\vartheta_j(s) - \vartheta_i(s)) \\
    &\qquad \qquad \qquad + {\beta}\Vert \bm{u}(s) \Vert_2^2 ds \;,
    \end{align*}
where the regularization term $\beta$.
  \begin{figure*}[t!]
        \subfloat[A complete graph with 64 nodes.]{%
            \includegraphics[width=.46\linewidth]{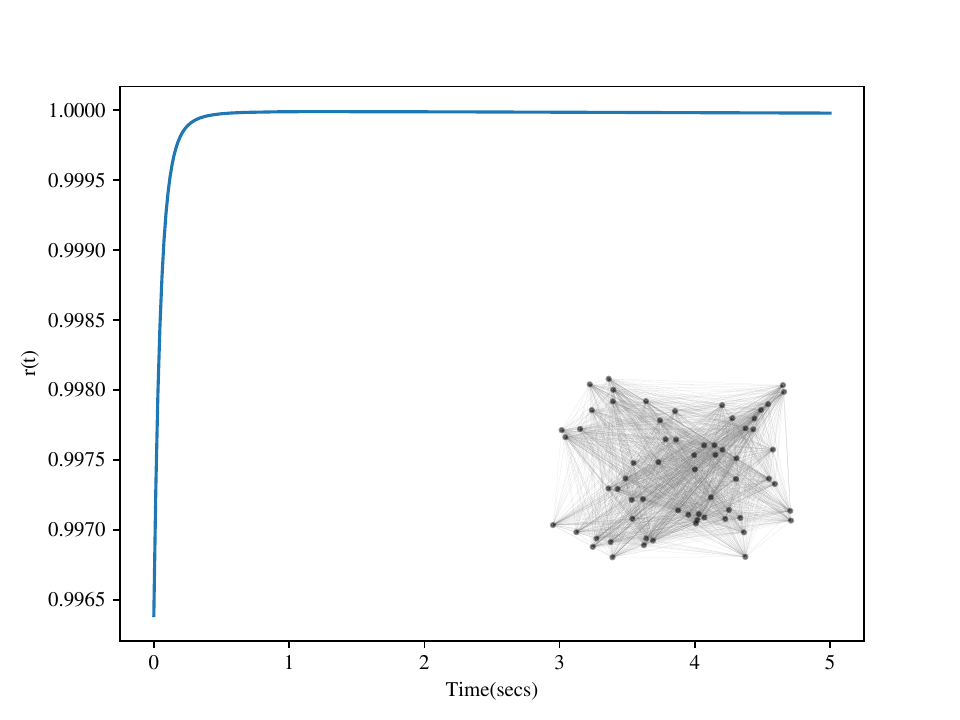}%
            \label{subfig:a}%
        }\hfill
        \subfloat[An Erdos-Renyi graph with 64 nodes.]{%
            \includegraphics[width=.46\linewidth]{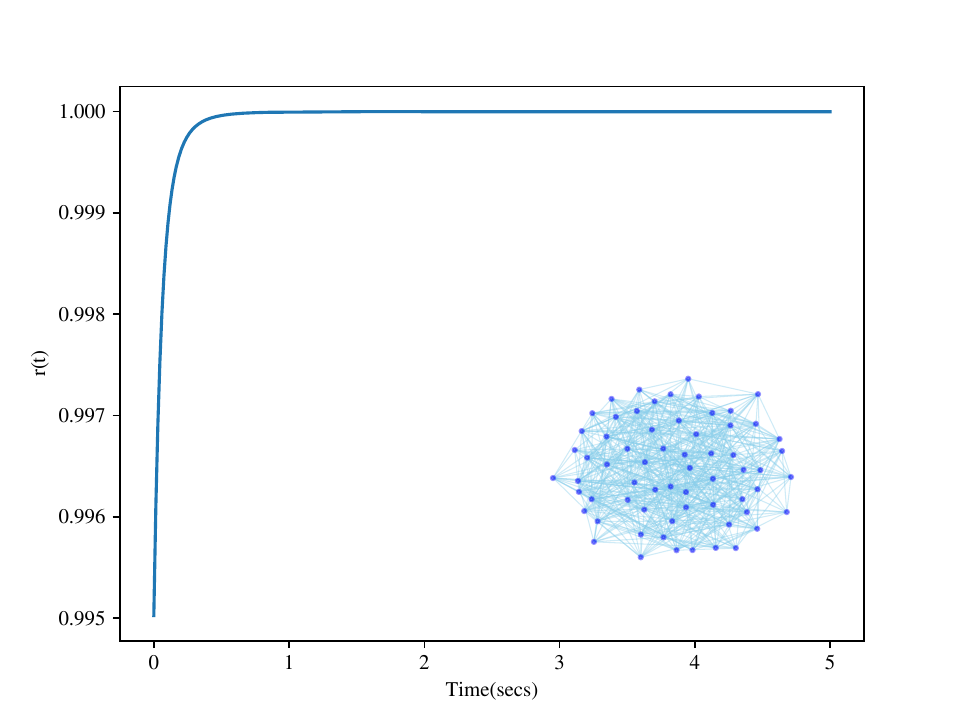}%
            \label{subfig:b}%
        }\\
        \subfloat[A Square Lattice with 8 by 8 grid.]{%
            \includegraphics[width=.46\linewidth]{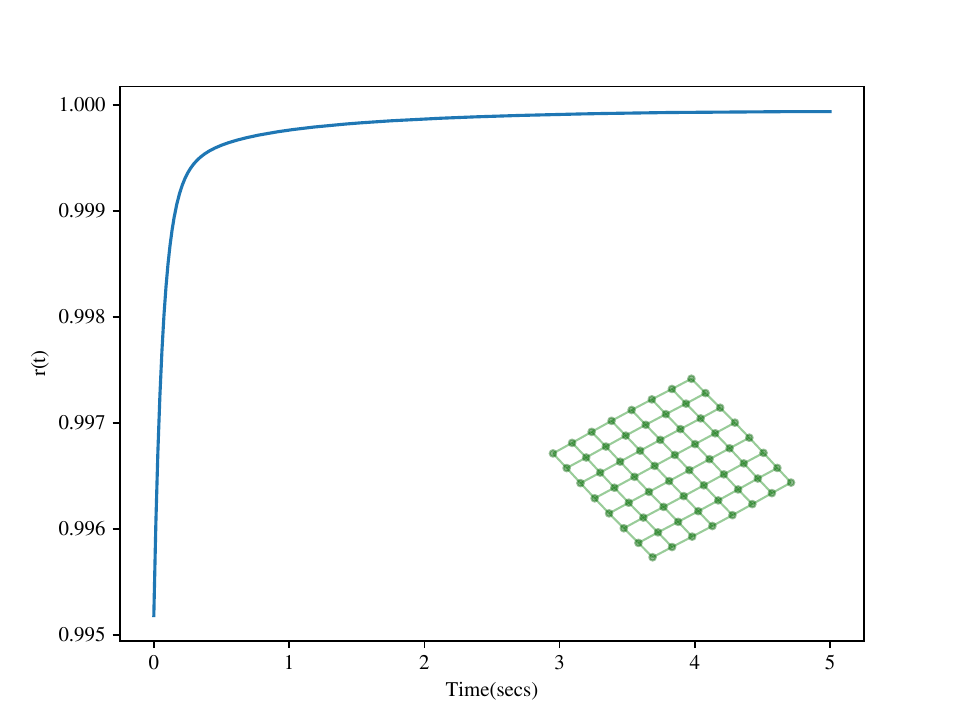}%
            \label{subfig:c}%
        }\hfill
        \subfloat[A Watt-Strogatz with 64 nodes.]{%
            \includegraphics[width=.46\linewidth]{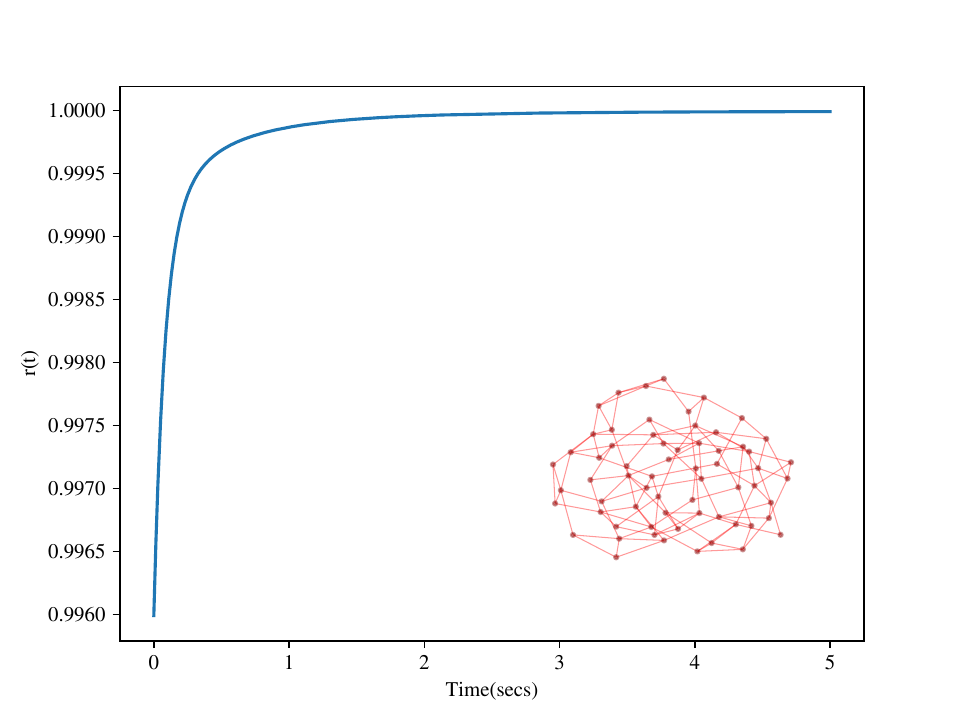}%
            \label{subfig:d}%
        }
        \caption{The consensus metric $r(t)$ for the closed-loop with different communication topologies exhibiting consensus.}
        \label{fig:kuramoto}
    \end{figure*}


We choose the parametrization of the neural pH controller \eqref{eq.Controller}
with trainable parameters of appropriate dimensions, $\bm{J}_c = \texttt{blkdiag}(J_i)$, $\bm{G}_c = \texttt{blkSparse}(\mathcal{P})$, and the Hamiltonian function $H_c$ as 
\begin{equation}
\label{eq:hamiltonian_example}
    H_c(\bm{\xi}) = \log \left( \cosh [ \texttt{blkdiag}(K_i) \bm{\xi} ] \right)^\top \mathds{1} \;,
\end{equation}
where $\mathds{1}$ is a vector of all ones.
Note that unlike RENs, where the storage function is always quadratic, our choice of Hamiltonian function is nonlinear. Moreover, one can also analytically compute the closed-form solution of the Jacobian of \eqref{eq:hamiltonian_example} for each sub-controller as 
\begin{equation*}
    \frac{\partial H_i}{\partial \xi_i}(\xi, \theta_i) = K^\top_i \tanh (K_i \xi_i), \ \forall i \in \mathcal{V}\;.
\end{equation*}

We trained the NN controller with the standard adjoint method and Forward Euler as the discretization scheme \cite{haber2017stable}.  The training is performed with  $500$ epochs using Adam \cite{kingma2015adam} with a learning rate of $5e-3$. We choose $\epsilon = 0.85$ and the finite horizon $T = 3.0$. To measure the degree of synchronization in the network, we introduce the metric 
\begin{gather*}
    r(t) := N^{-1} \sqrt{\sum_{i,j} \cos(\vartheta_{j}(t) - \vartheta_i(t))}, \ \forall i,j \in \mathcal{V} \;.
\end{gather*}
Note that a value of $r(t) = 1$ indicates that all oscillators have the same phase. 
The uncontrolled behaviour of oscillators under a fully-connected graph is plotted in Fig. \ref{fig:openloop}  and the closed-loop response of \eqref{eq:passive_kuramoto} and \eqref{eq.Controller} after the training is provided in Fig. \ref{fig:kuramoto}, respectively. 
Fig. \ref{fig:openloop} shows that the uncontrolled oscillators are not synchronized. 
On the other hand, Fig. \ref{fig:kuramoto} demonstrates the synchronization for different communication topologies. Particularly, we study the controller performance on a complete graph (grey lines, Fig. \ref{subfig:a}), an Erd\"{o}s–R\'{e}nyi network $\mathcal{G}(N,p)$ with $p=0.3$ (blue lines, Fig. \ref{subfig:b}), a square lattice (green lines, Fig. \ref{subfig:c}), and a Watts–Strogatz network with degree $k=5$ and a rewiring probability of 0.3 (red lines, Fig. \ref{subfig:d}) taken from \cite{asikis2022neural}. 
All networks consist of $N=64$ oscillators. 
 As shown, the NN controller effectively drives the oscillators to consensus (all $r(t)$ reach 1). Furthermore, we observe that consensus is maintained even after the finite-horizon $T=3$ used for optimization, demonstrating the closed-loop stability.\footnote{Our code is available at \href{https://github.com/DecodEPFL/Neural-Distributed-Controllers.git}{https://github.com/DecodEPFL/Neural-Distributed-Controllers.git}} 


\section{Conclusion} \label{sec:conclusion}
Neural distributed control of large-scale nonlinear systems can pose several challenges, such as guaranteeing closed-loop stability in an uncertain environment. 
To tackle this issue, we have proposed a free parametrization of neural distributed controllers via Hamiltonian structures that preserve closed-loop stability and guarantee a finite $\mathcal{L}_2$ gain, regardless of NN parameters, for arbitrarily large networks of nonlinear dissipative systems. We demonstarted that near-optimal performance can be achieved by parametrizing deep nonlinear storage functions for the controllers. Moreover, these NN structures can be leveraged for nonlinear system identification from data, where the identified neural models are stable by design and have a finite $\mathcal{L}_2$ gain. Further efforts will be devoted to exploring discretization schemes to preserve the $\mathcal{L}_2$ gain and, consequently, implementing the NN controllers on digital systems. 
\section*{Appendix}
\subsection{Proof of Theorem \ref{thm:finite_l2_gain}}
The proof is done by showing that the controller \eqref{eq.Controller}
is $\epsilon$-strictly output passive for $\alpha \geq \epsilon \bar{\lambda}(\bm{G}_c \bm{G}_c^\top)$. Recall from \cite{vanderSchaft2017}, that the controller \eqref{eq.Controller} is $\epsilon$-output strictly passive if and only if the following conditions hold $\forall \bm{\xi} \in \Xi$
\begin{subequations}
    \begin{equation} \label{eq.inq_1}
            \frac{\partial }{\partial \bm{\xi}}V f(\bm{\xi}) \leq  - \epsilon h^\top(\bm{\xi}) h(\bm{\xi})
    \end{equation}
    \begin{equation} \label{eq.inq_2}
        \frac{\partial }{\partial \bm{\xi}}V g(\bm{\xi}) = h^\top(\bm{\xi}) \;,  
    \end{equation}
\end{subequations}
where $V(\bm{\xi})$ is the $\mathcal{C}^1$ storage function and satisfies $V \geq 0$. 

Moreover, $f(\bm{\xi}) = (\bm{J} - \alpha \bm{I}) \frac{\partial H(\bm{\xi},\theta)}{\partial \bm{\xi}}$ for $\bm{\Lambda} = 0$, $g(\bm{\xi}) =  \bm{G}_c$, and $ h(\bm{\xi}) = \bm{G}^\top_c \frac{\partial H(\bm{\xi})}{\partial \bm{\xi}}$. 
First, we show that inequality \ref{eq.inq_1} is satisfied by design.
Let us choose the candidate storage function as the Hamiltonian of controller \eqref{eq.Controller}, i.e. $V(\bm{\xi}) = H_c(\bm{\xi}, \theta)$, then we have
\begin{align*}
     &\frac{\partial^\top H(\bm{\xi}) }{\partial \bm{\xi}} f(\bm{\xi}) \leq - \epsilon h^\top(\bm{\xi}) h(\bm{\xi}), \quad \forall \bm{\xi} \in \Xi  \\
      &\frac{\partial^\top H(\bm{\xi}) }{\partial \bm{\xi}} (\bm{J} - \alpha \bm{I}) \frac{\partial H(\bm{\xi})}{\partial \bm{\xi}} \leq - \epsilon  \frac{\partial^\top H(\bm{\xi}) }{\partial \bm{\xi}} \bm{G}_c \bm{G}^\top_c \frac{\partial H(\bm{\xi})}{\partial \bm{\xi}} \\
      &\frac{\partial^\top H(\bm{\xi}) }{\partial \bm{\xi}} \bigg( -\alpha \bm{I}  + \epsilon  \bm{G}_c \bm{G}^\top_c   \bigg) \frac{\partial H(\bm{\xi})}{\partial \bm{\xi}} \leq 0 \; .
\end{align*}
Therefore, choosing  $\alpha \geq \epsilon \bar{\lambda}(\bm{G}_c \bm{G}_c^\top)$ verifies the last inequality.
Note that the second equality \eqref{eq.inq_2} is verified by construction due to the choice of storage function and the structure of the port-Hamiltonian controller \eqref{eq.Controller}. 
Finally, by employing \cite[Theorem 2.2.13]{vanderSchaft2017} we conclude that if the controller \eqref{eq.Controller} is $\epsilon$-output strictly passive, then it has a finite $\mathcal{L}_2$-gain $ \leq 1/\epsilon$ for all trainable parameters $\bm{\theta}$. \hfill \QEDclosed




\bibliography{bibliography}
\bibliographystyle{IEEEtran}

\end{document}